\documentclass[%
 reprint,
superscriptaddress,
 amsmath,amssymb,
 aps,prb
]{revtex4-2}

\usepackage{graphicx}% Include figure files
\usepackage{dcolumn}% Align table columns on decimal point
\usepackage{bm}% bold math
\usepackage{hyperref}% add hypertext capabilities
\hypersetup{
	colorlinks   = true, %Colours links instead of ugly boxes
	urlcolor     = blue, %Colour for external hyperlinks
	linkcolor    = blue, %Colour of internal links
	citecolor    = blue %Colour of citations
}
\usepackage{pifont}
\usepackage{multirow}
\usepackage{tabularray}
\usepackage{array}
    \newcolumntype{P}[1]{>{\centering\arraybackslash}p{#1}}
    \newcolumntype{M}[1]{>{\centering\arraybackslash}m{#1}}
\usepackage{stackengine}
    
\usepackage{romannum}
\usepackage{bigints}
\usepackage{ulem}
\usepackage{xcolor}
\usepackage{tabularx} % table related
\usepackage{siunitx} % table related
\usepackage[footnotehyper]{nicematrix}
\NiceMatrixOptions{cell-space-limits = 1.4pt}

\begin{document}

\preprint{APS/123-QED}

\title{Superconducting coherence peak in near-field radiative heat transfer}

\author{Wenbo Sun}
\affiliation{Elmore Family School of Electrical and Computer Engineering, Birck Nanotechnology Center, Purdue University, West Lafayette, Indiana 47907, USA}
\author{Zhuomin M. Zhang}
\affiliation{George W. Woodruff School of Mechanical Engineering, Georgia Institute of Technology, Atlanta, Georgia 30332, USA}
\author{Zubin Jacob}
\email{zjacob@purdue.edu}
\affiliation{Elmore Family School of Electrical and Computer Engineering, Birck Nanotechnology Center, Purdue University, West Lafayette, Indiana 47907, USA}

\date{\today}

\begin{abstract}
Enhancement and peaks in near-field radiative heat transfer (NFRHT) typically arise due to surface phonon-polaritons, plasmon-polaritons, and electromagnetic (EM) modes in structured materials. However, the role of material quantum coherence in enhancing near-field radiative heat transfer remains unexplored. Here, we unravel that NFRHT in superconductor-ferromagnetic systems displays a unique peak at the superconducting phase transition that originates from the quantum coherence of Bogoliubov quasiparticles in superconductors. Our theory takes into account evanescent EM radiation emanating from fluctuating currents related to Cooper pairs and Bogoliubov quasiparticles in stark contrast to the current-current correlations induced by free electrons in conventional materials. Our proposed NFRHT configuration exploits ferromagnetic resonance at frequencies deep inside the superconducting band gap to isolate this superconducting coherence peak. Furthermore, we reveal that Cooper pairs and Bogoliubov quasiparticles have opposite effects on near-field thermal radiation and isolate their effects on many-body radiative heat transfer near superconductors. Our proposed phenomenon can have applications for developing thermal isolators and heat sinks in superconducting circuits.
\end{abstract}

\maketitle

\normalem

\section{Introduction}
Near-field radiative heat transfer (NFRHT) is mediated by evanescent electromagnetic (EM) fields emitted from fluctuating current sources inside materials~\cite{biehs2021near,volokitin2007near,zhang2007nano}, and plays a crucial role in thermal energy technologies. Considerable efforts have focused on enhancing NFRHT by using plasmons, phonons, and photonic modes in metals~\cite{salihoglu2023nonlocal,volokitin2002dissipative,hassani2022drifting,ilic2018active,tokunaga2021extreme,tan2016enhancing,kallel2017temperature}, polar dielectrics~\cite{feng2024near,sasihithlu2018dynamic,feng2023phonon}, hyperbolic materials~\cite{guo2012broadband,feng2022thermoradiative}, and structured materials~\cite{venkataram2020fundamental,zhao2017near,luo2024near,asheichyk2024heat,nolen2024local,huang2023temporal}, where the fluctuating currents are generally related to thermal motions of electrons or ions~\cite{greffet2007coherent}. Recent research interests focused on exploring new material platforms for controlling nanoscale radiative thermal transport~\cite{yu2023manipulating,zhou2024unconventional,kralik2017effect}. Special interest is given to phase-transition materials with more ordered electronic phases, e.g., superconductors~\cite{musilova2019strong,kralik2017effect,moncada2021normal,castillo2022enhancing} or charge density waves~\cite{zhou2024unconventional}. Near-field thermal radiation from these new material platforms not only exhibits unconventional behaviors important for developing energy harvesting devices, but also encodes important material information useful for building novel probes of materials~\cite{degottardi2019thermal,zhou2024unconventional}. 

One fundamental difference between thermal radiation from superconductors and metals/polar dielectrics is that fluctuating currents in superconductors are related to Cooper pairs and Bogoliubov quasiparticles (broken Cooper pairs) rather than free electrons/ions~\cite{dressel2013electrodynamics}. Distinct from free electrons in normal conductors, Cooper pairs and Bogoliubov quasiparticles exhibit intrinsic quantum coherence effects~\cite{coleman2015introduction}. As an example, Cooper pairs consist of two electrons with opposite spins, while Bogoliubov quasiparticles are coherent superpositions of electrons and holes. This superconducting coherence reveals the internal structures of the BCS condensate and is crucial for understanding superconductivity. Previous works focused on revealing the superconducting coherence effects in electrical transport~\cite{klein1994conductivity,boyack2023electrical}, quasiparticle scattering~\cite{hanaguri2009coherence,zou2022particle}, and nuclear spin relaxation~\cite{hebel1959nuclear,dai2024existence,li2024unconventional}. However, the role of superconducting coherence in NFRHT remains unexplored.

\begin{figure}[!t]
    \centering
    \includegraphics[width= 3in]{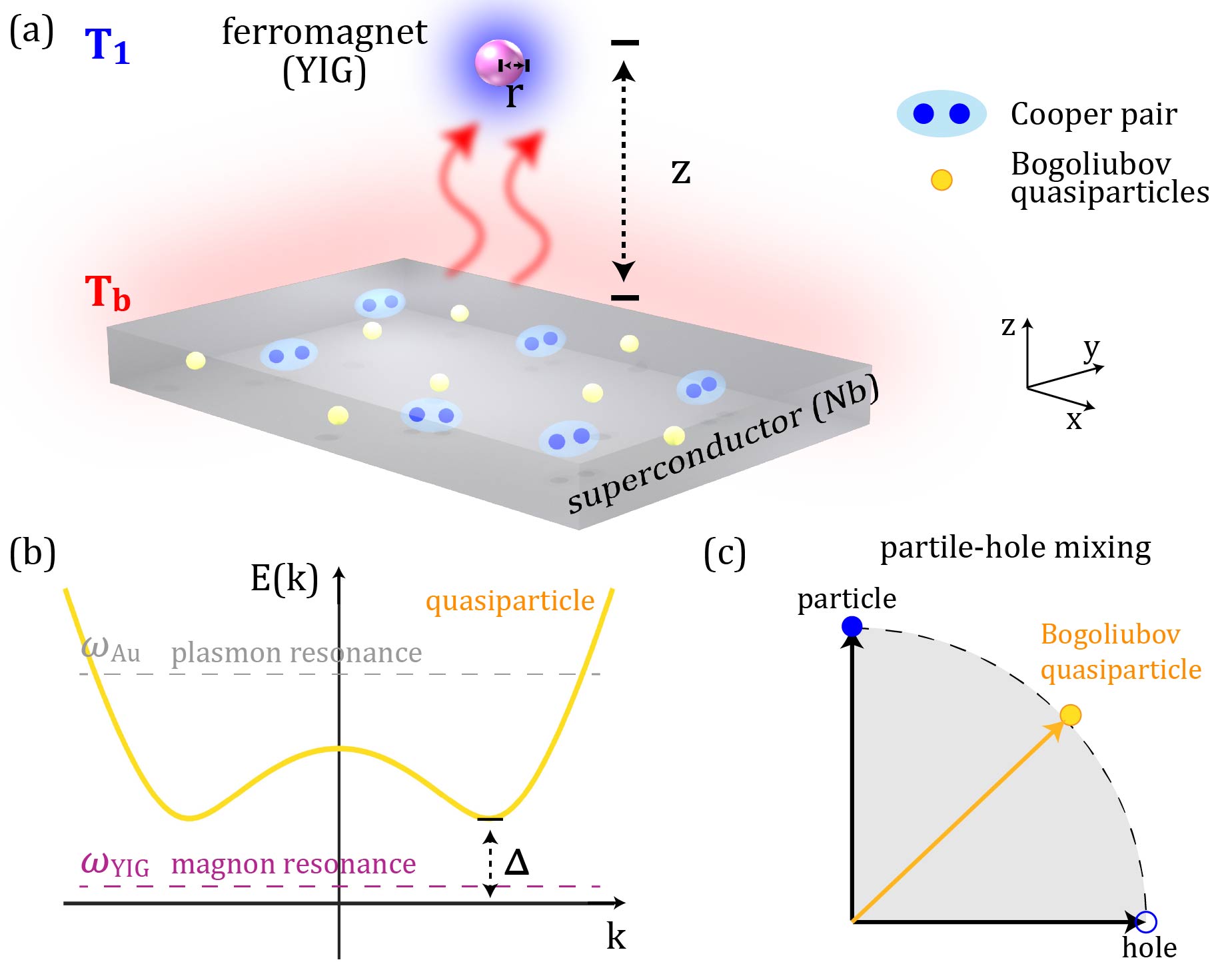}
    \caption{(a) Design of the superconductor-ferromagnet system to unravel superconducting coherence effects in thermal photonics. (b) Energy spectra of Bogoliubov quasiparticles and superconducting band gap $\Delta$. We exploit low-frequency YIG magnon resonance $\omega_\mathrm{YIG}$ deep inside the superconducting band gap to unravel the superconducting coherence effects in NFRHT. (c) Bogoliubov quasiparticles are coherent quantum superpositions of electrons and holes, known as electron-hole mixing.}
    \label{fig:fig1}
\end{figure}

Previous works about NFRHT and superconductivity focused on dielectric/normal-metal-superconductor systems and found that NFRHT is strongly suppressed by superconductivity~\cite{musilova2019strong,kralik2017effect,moncada2021normal,castillo2022enhancing} at the superconducting phase transition. In those systems, thermal radiation is dominated by spectrum components at frequencies comparable to or higher than the superconducting band gap~\cite{moncada2021normal,castillo2022enhancing} (Fig.~\ref{fig:fig1}(b)). Meanwhile, superconducting coherence effects are usually connected to low-frequency response deep inside the superconducting band gap. This indicates the necessity of exploring systems with NFRHT dominated by low-frequency spectrum components to unravel superconducting coherence effects in thermal photonics.

In this paper, we exploit the low-frequency ferromagnetic resonance in yttrium iron garnet (YIG) to reveal superconducting coherence effects in NFRHT between YIG nanoparticles and a superconducting niobium slab (Fig.~\ref{fig:fig1}(a)). We demonstrate that NFRHT in the superconductor-ferromagnet system exhibits distinct behaviors compared to superconductor-normal-metal/dielectric systems. Specifically, we find a superconducting coherence peak that enhances NFRHT at the phase transition, which serves as the fingerprint of superconducting coherence effects. This superconducting coherence peak leads to a negative temperature dependence of NFRHT, which is different from conventional NFRHT behaviors between blackbodies or common materials. In addition, we reveal that Cooper pairs and Bogoliubov quasiparticles have opposite effects on NFRHT in this superconductor-ferromagnetic system, which can be exploited to build novel nanoscale thermal devices. Finally, we isolate the effects of Bogoliubov quasiparticles and Cooper pairs on the spatial coherence of near-field thermal radiation, and their effects on many-body NFRHT in YIG nanoparticle arrays near a superconducting slab. 

\section{Quantum coherence and Bogoliubov quasiparticles}
In superconductors, attractive interactions (e.g., phonon-mediated interactions) between electrons open a superconducting band gap $\Delta$ at the Fermi surface, leading to the formation of two unique types of quasiparticles, i.e., Cooper pairs and Bogoliubov quasiparticles~\cite{dressel2013electrodynamics}. As shown in Fig.~\ref{fig:fig1}(b), Cooper pairs are paired electrons that condense into the ground state, while Bogoliubov quasiparticles are excitations (broken Cooper pairs) separated by the superconducting band gap $\Delta$ from the ground state~\cite{coleman2015introduction}. 

One unique signature of Bogoliubov quasiparticles in superconductors is their quantum nature as coherent superpositions of electrons and holes~\cite{campuzano1996direct,fujita2008bogoliubov}  (see Fig.~\ref{fig:fig1}(c)). This is explicitly demonstrated by the creation operator of Bogoliubov quasiparticles $a^{\dagger}_{\textbf{k},\sigma}$~\cite{coleman2015introduction},
\begin{equation}
    a^{\dagger}_{\textbf{k},\sigma} = u_{\textbf{k}} c^{\dagger}_{\textbf{k},\sigma} + v_{\textbf{k}} \mathrm{sgn}(\sigma) c_{-\textbf{k},-\sigma},
\end{equation}
where $c^{\dagger}_{\textbf{k},\sigma}$ represents the creation of an electron with momentum $\textbf{k}$ and spin $\sigma$, and $c_{-\textbf{k},-\sigma}$ represents the creation of a hole (annihilation of an electron) with momentum $-\textbf{k}$ and spin $-\sigma$. $u_{\textbf{k}}, v_{\textbf{k}}$ are the coefficients in the Bogoliubov transformation satisfying $|u_{\textbf{k}}|^2 + |v_{\textbf{k}}|^2 = 1$, which represent the coherent superposition of electrons and holes for Bogoliubov quasiparticles. These coefficients $u_{\textbf{k}}, v_{\textbf{k}}$ (sometimes also termed coherence factors) reveal the internal structures of condensed electron pairs~\cite{hanaguri2009coherence}, and manifest themselves in EM response only at frequencies deep inside the superconducting band gap $\omega \ll \omega_g=2\Delta/\hbar$~\cite{klein1994conductivity}. Previous works focused on unraveling these coherence factors in electrical transport~\cite{klein1994conductivity,boyack2023electrical}, quasiparticle scattering~\cite{hanaguri2009coherence,zou2022particle}, and nuclear spin relaxation~\cite{hebel1959nuclear,dai2024existence,li2024unconventional}. Our central goal in this paper is to reveal the fingerprints of the superconducting coherence factors in thermal photonics, which can also open new frontiers in manipulating radiative heat transfer in energy harvesting devices with superconducting components.

\begin{figure*}[!t]
    \centering
    \includegraphics[width= 5.6 in]{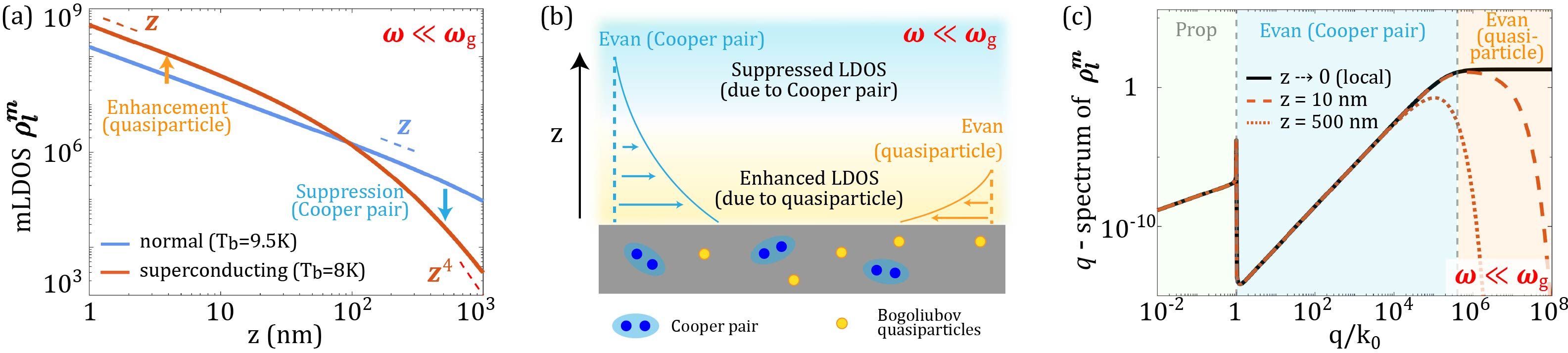}
    \caption{Opposing effects of Cooper pairs and Bogoliubov quasiparticles on low-frequency near-field thermal radiation. (a) Dependence of low-frequency magnetic LDOS $\rho^m_{l}$ on distance $z$ from the niobium slab. In the superconducting phase, $\rho^m_{l}$ is enhanced by Bogoliubov quasiparticle coherence at small distance, while suppressed by Cooper pairs screening effects at larger distance. (b) Sketch of evanescent waves associated with Bogoliubov quasiparticles and Cooper pairs. (c) Momentum $q$-spectrum of mLDOS $\rho^m_{l}$ near a superconducting niobium slab at $T=8\,\mathrm{K}$. High-momentum evanescent waves are dominated by Bogoliubov quasiparticles, while low-momentum evanescent waves are dominated by Cooper pair screening effects. The black line corresponds to the $q$-spectrum of $\rho^m_{l}$ for $z\rightarrow0$ in the local response limit (neglecting nonlocal effects). We consider frequency $\omega= 2\pi \, \mathrm{GHz} \ll \omega_g=2\Delta/\hbar \approx 4\,\mathrm{THz}$ in this figure.}
    \label{fig:fig2}
\end{figure*}

\section{Model}
As shown in Fig.~\ref{fig:fig1}(a), we consider a spherical gyromagnetic YIG nanoparticle with radius $r$ at temperature $T_1$ placed near a niobium slab at temperature $T_b$. We assume that the particle is small compared to its distance $z$ from the slab $r < z$ and consider the dipole approximation for the YIG nanoparticle. Following the framework of fluctuational electrodynamics~\cite{messina2013fluctuation}, we find the net power $P_1$ absorbed by the YIG nanoparticle is (see derivations in Appendix~\ref{AppA}),
\begin{align}\label{P1}
\begin{aligned}
    P_1 =8 \int_0^\infty d\omega \hbar \omega k_0^2 (n_b-n_1) \, \mathrm{Tr} \{ \mathrm{Im}[\Bar{\Bar{G}}_m(\Vec{r}_1,\Vec{r}_1)] \mathrm{Im}[\Bar{\Bar{\alpha}}_1] \},
\end{aligned}
\end{align}
where $k_0=\omega/c$ is the vacuum wavenumber, $n_i=1/(e^{\hbar\omega/k_B T_i}-1)$ is the mean photon number at temperature $T_i$ ($i=1,b$), $\Bar{\Bar{G}}_m(\Vec{r}_1,\Vec{r}_1)$ is the near-field magnetic dyadic Green's function at the nanoparticle position $\Vec{r}_1$, and $\Bar{\Bar{\alpha}}_1$ is the magnetic polarizability tensor of the YIG nanoparticle. As shown in Fig.~\ref{fig:fig1}(b), YIG typically has magnon resonance at GHz frequencies $\omega_{\mathrm{YIG}}$ much lower than the superconducting band gap~\cite{khosravi2024giant}. This leads to a peak of $\mathrm{Im} [\Bar{\Bar{\alpha}}_1]$ and spectral heat transfer at $\omega_{\mathrm{YIG}} \ll \omega_g$~\cite{khosravi2024giant}. Therefore, from Eq.~(\ref{P1}), NFRHT in our system is governed by low-frequency spectrum components deep inside the superconducting band gap, allowing the isolation of superconducting coherence effects in this ferromagnet-superconductor system. This is in stark contrast to NFRHT in previous metal/dielectric-superconductor systems, which is generally determined by spectrum components larger than or comparable to the superconducting band gap~\cite{moncada2021normal,musilova2019strong,kralik2017effect}. We further note that plasmon resonance in conventional metals or dielectrics is typically at frequencies much higher than the superconducting band gap and cannot be exploited to isolate superconducting coherence effects in thermal photonics (see Fig.~\ref{fig:fig1}(b)).

From Eq.~(\ref{P1}), effects of Cooper pairs and Bogoliubov quasiparticles on NFRHT are encoded in the near-field magnetic Green's function $\Bar{\Bar{G}}_m$. Here, in the near-field regime $z\ll k_0^{-1}$, $\Bar{\Bar{G}}_m$ is dominated by s-polarized evanescent waves~\cite{joulain2003definition,chapuis2008effects,zhang2007nano,volokitin2007near} and can be approximated as (see Appendix~\ref{AppB}),
\begin{equation}\label{dyadG}
    \Bar{\Bar{G}}_m (\Vec{r}_1,\Vec{r}_1) \approx \frac{i}{8 \pi k_0^2}\int \frac{q dq}{k_z} e^{-2q z} r_s(q) \begin{bmatrix} -k_z^2 & 0 & 0\\0 & -k_z^2 & 0\\0 & 0 & 2q^2 \end{bmatrix},
\end{equation}
where $k_z=\sqrt{k_0^2-q^2}$, $q$ is the in-plane wavevector of evanescent waves, and $z$ is the distance between $\Vec{r}_1$ and the slab interface. The Fresnel reflection coefficient $r_s(q)$ is determined by the permittivity of the niobium slab $\varepsilon_{\mathrm{Nb}}$. In this paper, we consider the Drude model for $\varepsilon_{\mathrm{Nb}}$ in the normal phase, and Mattis-Bardeen theory~\cite{mattis1958theory,popel1989surface} for $\varepsilon_{\mathrm{Nb}}$ in the superconducting phase (see details in Appendix~\ref{AppC}). In the superconducting phase, Bogoliubov quasiparticles contribute to the dissipation of the superconductor, i.e., $\mathrm{Im} \, \varepsilon_{\mathrm{Nb}}$. Different from normal conduction electrons, the coherence of Bogoliubov quasiparticles introduces an additional coherence factor, 
\begin{equation}\label{coh_factor}
    u_{\textbf{k}}u_{\textbf{k}'} + v_{\textbf{k}}v_{\textbf{k}'},
\end{equation}
that renormalizes the scattering of quasiparticles from the state $\textbf{k}$ to another state $\textbf{k}'$ upon absorption of a photon with momentum $\textbf{k}'-\textbf{k}$~\cite{coleman2015introduction,dressel2013electrodynamics}. At low frequencies $\omega \ll \omega_g$, this additional coherence factor becomes prominent and leads to increased dissipation in the superconducting phase~\cite{klein1994conductivity}. In contrast, Cooper pairs contribute to the screening effects (negative $\mathrm{Re} \, \varepsilon_{\mathrm{Nb}}$) much stronger compared to the screening effects in the normal phase. 

\section{Near-field magnetic local density of states}\label{sec_mldos}
Our first goal is to isolate different effects of Cooper pairs and Bogoliubov quasiparticles on low-frequency near-field thermal radiation, which can be characterized by photonic local density of states (LDOS). In Fig.~\ref{fig:fig2}(a), we study low-frequency magnetic LDOS (mLDOS) $\rho^m_{l}=\omega \mathrm{Tr} \{ \mathrm{Im}[\Bar{\Bar{G}}_m(\Vec{r}_1,\Vec{r}_1)] \} / \pi c^2$ at $1\,\mathrm{GHz}$ near a superconducting niobium slab (transition temperature $T_c=9.2\,\mathrm{K}$). We find that, right below the transition temperature $T_c$, low-frequency $\rho_m$ is enhanced by around 3 times at small distance ($z < 100\,\mathrm{nm}$) from the superconducting slab, but is strongly suppressed at relatively larger distance ($z \geq 100\,\mathrm{nm}$).  

\begin{figure*}[!t]
    \centering
    \includegraphics[width=6 in]{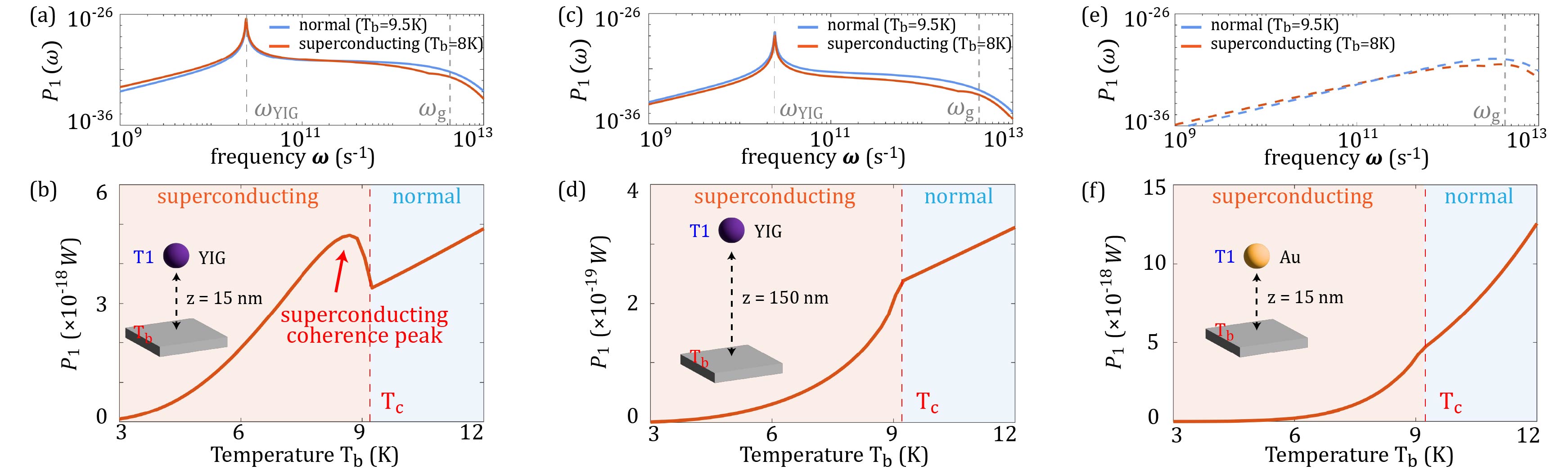}
    \caption{Superconducting coherence peak in NFRHT due to Bogoliubov quasiparticle coherence. (a-d) Radiative heat transfer $P_1$ and spectral heat transfer $P_1(\omega)$ between a YIG nanoparticle and a niobium slab separated by (a,b) $z=15\,\mathrm{nm}$ or (c,d) $z=150\,\mathrm{nm}$. (e,f) $P_1$ and $P_1(\omega)$ between a gold nanoparticle and a niobium slab separated by $z=15\,\mathrm{nm}$. Quantum coherence of Bogoliubov quasiparticles leads to the peak of $P_1$ right below the transition temperature in (b). We consider $r=6\, \mathrm{nm}$ and $T_1=1 \, \mathrm{K}$ in this figure. Detailed material parameters are provided in Appendix~\ref{AppC}.} 
    \label{fig:fig3}
\end{figure*} 

As sketched in Fig.~\ref{fig:fig2}(b), the different behaviors of $\rho^m_{l}$ in the small and large $z$ regions at the superconducting phase transition originate from evanescent waves associated with different mechanisms. To elucidate this idea, we analyze the momentum $q$-spectrum of near-field mLDOS $\partial \rho^m_{l} / \partial q$ when the niobium slab is in the superconducting phase. As shown in Fig.~\ref{fig:fig2}(c), the $q$-spectrum can be divided into three regions corresponding to different mechanisms. At $q < k_0$, $k_z=\sqrt{k_0^2-q^2}$ is real, indicating contributions from propagating waves. Meanwhile, at $q > k_0$, $e^{-2qz}$ in Eq.~(\ref{dyadG}) provides a momentum cut-off at $q \sim 1/2z$. Therefore, in the small $z$ region, mLDOS $\rho^m_{l}$ is dominated by contributions from high-momentum evanescent waves with $q\gg k_0$. In this region, we have,
\begin{equation}\label{smallz}
    \partial\rho^m_{l} / \partial q \approx \frac{\omega}{8\pi^2c^2} \mathrm{Im} [\varepsilon_\mathrm{Nb}-1] e^{-2qz}.
\end{equation}
We thus ascribe these high-momentum evanescent waves to dissipation associated with Bogoliubov quasiparticles. In this small $z$ region, quantum coherence of Bogoliubov quasiparticles enhances the mLDOS $\rho^m_{l}$ at the superconducting phase transition. Integrating Eq.~(\ref{smallz}) with respect to $q$, we find $\rho^m_{l} \sim z^{-1}$ matching our numeric calculations in Fig.~\ref{fig:fig2}(a).

At relatively larger $z$, magnetic LDOS $\rho^m_{l}$ is dominated by contributions from evanescent waves with lower momentum. In this region, we have,
\begin{equation}\label{largez}
    \partial\rho^m_{l} / \partial q \approx \frac{c q^3}{\pi^2\omega^2} \mathrm{Im} [\frac{i}{\sqrt{\varepsilon_\mathrm{Nb}}}] e^{-2qz}.
\end{equation}
In the superconducting phase, $|\mathrm{Re} [\varepsilon_\mathrm{Nb}]| \gg \mathrm{Im} [\varepsilon_\mathrm{Nb}]$ at low frequencies. Therefore, we attribute these lower-momentum evanescent waves to be dominated by Cooper pairs. In this large $z$ region, Cooper pair screening effects strongly suppress $\rho^m_{l}$ at the superconducting phase transition. Integrating Eq.~(\ref{largez}), we find the mLDOS $\rho^m_{l} \sim z^{-4}$ matching our numeric calculations in Fig.~\ref{fig:fig2}(a).

\section{Superconducting coherence peak in near-field radiative heat transfer}
We now reveal the signatures of Bogoliubov quasiparticle coherence in NFRHT. Evanescent waves associated with Bogoliubov quasiparticles and Cooper pairs lead to distinct NFRHT behaviors in the small and large $z$ regions at the superconducting phase transition. Here, we consider NFRHT $P_1$ between a YIG nanoparticle at temperature $T_1=1\,\mathrm{K}$ and a niobium slab at temperature $T_b$ separated by $z=15\,\mathrm{nm}$ (Figs.~\ref{fig:fig3}(a,b)) and $z=150\,\mathrm{nm}$ (Figs.~\ref{fig:fig3}(c,d)). As shown in Figs.~\ref{fig:fig3}(a,c), YIG magnon resonance leads to a peak in spectral heat transfer $P_1(\omega)$ at $\omega_{\mathrm{YIG}}$. Therefore, NFRHT in this superconductor-ferromagnet system is dominated by low-frequency spectrum components at $\omega \ll \omega_g$. 

For a small vacuum gap $z=15\,\mathrm{nm}$, NFRHT is dominated by evanescent waves associated with Bogoliubov quasiparticles in the superconducting phase. At low frequencies $\omega \ll \omega_g$, Bogoliubov quasiparticles contribute to increased EM dissipation in the superconducting phase right below $T_c$ compared to the normal phase. This increased dissipation originates from the coherence factor Eq.~(\ref{coh_factor}), which stems from the coherence of Bogoliubov quasiparticles and becomes prominent at low frequencies. As discussed in Sec.~\ref{sec_mldos}, this increased EM dissipation then enhances the mLDOS at $\omega \ll \omega_g$ in the small $z$ region. Therefore, as shown in Fig.~\ref{fig:fig3}(b), we find that the heat transfer $P_1$ is enhanced right below the transition temperature $T_c$ due to the quantum coherence of Bogoliubov quasiparticles. This reveals a superconducting `coherence peak' in near-field thermal conductance, which is the fingerprint of superconductor coherence factors in thermal photonics. This peak leads to an unconventional negative temperature dependence of heat transfer $P_1$ at the superconducting phase transition, which purely originates from the superconducting coherence effects and is in stark contrast to the positive temperature dependence of blackbody radiation or radiation from common materials. Meanwhile, we find that $P_1$ decays fast at lower temperatures due to a reduced number of thermally excited Bogoliubov quasiparticles. We note that the above NFRHT behaviors in this superconductor-ferromagnet system are in stark contrast to previous NFRHT observed in superconductor-normal-metal systems~\cite{musilova2019strong,kralik2017effect} (also see Fig.~\ref{fig:fig3}(f)).

In contrast, for a relatively larger vacuum gap $z=150\,\mathrm{nm}$, we find the heat transfer $P_1$ is strongly suppressed below the transition temperature $T_c$. This strong suppression originates from near-field thermal radiation dominated by Cooper pair screening effects. As discussed in Sec.~\ref{sec_mldos}, Cooper pairs contribute to strong screening effects in the superconducting phase, which suppress the mLDOS in the large $z$ region. We note that, in this case, $P_1$ decreases monotonically with decreasing temperature, and the effects of superconducting coherence factors cannot be observed.

For comparison, in Figs.~\ref{fig:fig3}(e,f), we also demonstrate NFRHT between a gold nanoparticle and a niobium slab separated by $z=15\,\mathrm{nm}$. As shown in Fig.~\ref{fig:fig3}(e), in stark contrast to YIG nanoparticles, NFRHT in this superconductor-normal-metal system is dominated by spectral components at frequencies $\omega \gtrsim \omega_g$. At high frequencies, the coherence factor becomes less appreciable~\cite{dressel2013electrodynamics}, and the EM dissipation is suppressed compared to the normal state and monotonically decreases with temperature~\cite{klein1994conductivity}. This decreased EM dissipation then suppresses the mLDOS at $\omega \gtrsim \omega_g$ in the small $z$ region, in stark contrast to the $\omega \ll \omega_g$ frequency regime in Fig.~\ref{fig:fig2}(a). Therefore, as shown in Fig.~\ref{fig:fig3}(f), $P_1$ decreases monotonically with decreasing temperature, and no superconducting `coherence peak' is revealed in NFRHT. 

\section{Superconductivity effects on many-body NFRHT}
\begin{figure}[!t]
    \centering
    \includegraphics[width=3.2 in]{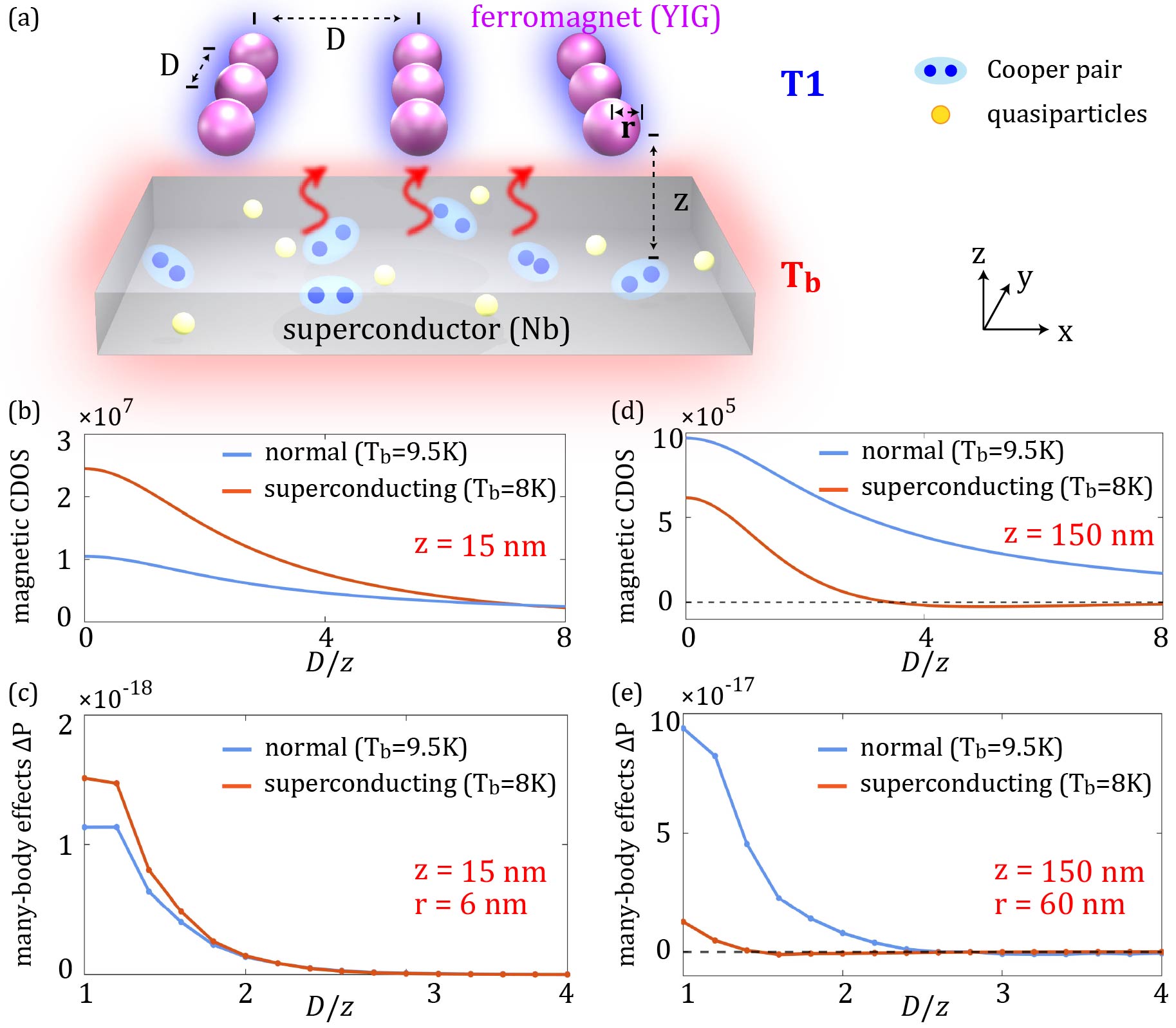}
    \caption{Isolating effects of Cooper pairs and Bogoliubov quasiparticles on many-body NFRHT. (a) Schematic of NFRHT between a $3 \times 3$ YIG nanoparticle array at temperature $T_1$ and a niobium slab at temperature $T_b$. Nanoparticles are separated from the niobium slab by a distance $z$. (b,c) Magnetic cross density of states $\rho_c^m$ at $\omega=2\pi \mathrm{GHz}$ at two points separated by distance $D$ at the same distance z from the niobium slab. (d,e) Many-body effects $\Delta P = P_{mb}-P_1$ on NFRHT when the niobium slab is in the normal phase or superconducting phase. Parameters are (d) $T_1=1 \, \mathrm{K}, r=6\, \mathrm{nm}, P_1(9.5\,\mathrm{K}) \approx 3.5\times 10^{-18} \,\mathrm{W}, P_1(8\,\mathrm{K}) \approx 4.3\times 10^{-18} \,\mathrm{W}$, and (e) $T_1=1 \, \mathrm{K}, r=60\, \mathrm{nm}, P_1(9.5\,\mathrm{K}) \approx 2.5\times 10^{-16} \,\mathrm{W}, P_1(8\,\mathrm{K}) \approx 1.1\times 10^{-16} \,\mathrm{W}$.}
    \label{fig:fig4}
\end{figure}

We have isolated the effects of Cooper pairs and Bogoliubov quasiparticles on magnetic LDOS and NFRHT between two bodies. Meanwhile, another important property of near-field thermal radiation is its spatial coherence~\cite{yu2023manipulating,bailly2021spatial}, which is crucial for many-body NFRHT~\cite{zhu2018theory,messina2013fluctuation}. As shown in Fig.~\ref{fig:fig4}(a), we now consider a $3 \times 3$ array of identical YIG nanoparticles with polarizability tensor $\Bar{\Bar{\alpha}}_1$ at temperature $T_1$ near a niobium slab at temperature $T_b$. We consider that all particles are maintained at the same temperature $T_1$, thus there is no net heat transfer between the particles. We focus on the net thermal power $P_{mb}$ absorbed by the central particle at position $\Vec{r}_{1}$ in the array. Following the framework of fluctuational electrodynamics~\cite{messina2013fluctuation}, we find (see derivations in Appendix~\ref{AppA}),
\begin{multline}\label{many-body}
    P_{mb}=8 \int_0^\infty  d\omega \hbar \omega k_0^2 (n_b-n_1) \, \mathrm{Im} \big\{ \sum_{jj'}^{N=9} \mathrm{Tr} \big[ [\Bar{\Bar{T}}^{-1}]_{1j} \\ 
    [\Bar{\Bar{A}}]_{jj} \mathrm{Im} [\Bar{\Bar{G}}]_{jj'} [\Bar{\Bar{A}}^{\dagger}]_{j'j'} [\Bar{\Bar{T}}^{-1,\dagger}]_{j'1} [\Bar{\Bar{B}}^{\dagger}]_{11} \big] \big\},
\end{multline}
where $ \Bar{\Bar{T}}_{ij} = \delta_{ij} \Bar{\Bar{I}} - 4\pi k_0^2 (1-\delta_{ij}) \Bar{\Bar{\alpha}}_1 \Bar{\Bar{G}}_{ij}$, $\Bar{\Bar{A}}_{ij} = \delta_{ij} \Bar{\Bar{\alpha}}_1$, $\Bar{\Bar{B}}_{ij} = \delta_{ij} (\Bar{\Bar{\alpha}}_1^{-1} + 4\pi k_0^2 \Bar{\Bar{G}}_{ij})$, and $\Bar{\Bar{G}}_{ij} = \Bar{\Bar{G}}(\Vec{r}_i,\Vec{r}_{j})$. $\Bar{\Bar{I}}$ is the $3 \times 3$ identity matrix and $\delta_{ij}$ is the Kronecker delta. We characterize the many-body effects on NFRHT by $\Delta P = P_{mb} - P_1$, which represents the differences of net power absorbed by the YIG nanoparticle with and without the presence of other nanoparticles.

From Eq.~(\ref{many-body}), we note that many-body effects $\Delta P$ are largely determined by the spatial coherence of near-field thermal fluctuations proportional to $\mathrm{Im} [\Bar{\Bar{G}}]_{jj'}$ in our system. Therefore, in Figs.~\ref{fig:fig4}(b,d), we first study the effects of superconductivity on the magnetic cross density of states (mCDOS) $\rho^m_c = \omega \mathrm{Tr} \{ \mathrm{Im}[\Bar{\Bar{G}}_m(\Vec{r}_1,\Vec{r}_2)] \} / \pi c^2$, which characterize the correlations of thermal fluctuations at two points $\Vec{r}_1$ and $\Vec{r}_2$. We focus on $\rho^m_c$ at two points separated by distance $D=|\Vec{r}_1 - \Vec{r}_2|$ at the same distance $z$ from the niobium slab. Similar to mLDOS, evanescent waves associated with Cooper pairs and Bogoliubov quasiparticles lead to distinct behaviors of mCDOS in the small and large $z$ regions at the superconducting phase transition. As shown in Fig.~\ref{fig:fig4}(b), in the small $z=15\,\mathrm{nm}$ region, we find that mCDOS at two points separated by small distance $D$ are enhanced by quantum coherence of Bogoliubov quasiparticles. Meanwhile, at larger separation $D$, mCDOS are suppressed due to Cooper pair screening effects. In contrast, as shown in Fig.~\ref{fig:fig4}(d), we find mCDOS at large $z=150\,\mathrm{nm}$ are strongly suppressed at all $D$ due to Cooper pair screening effects. In general, we find that the strong EM screening effects from Cooper pairs can suppress the spatial coherence length of near-field thermal radiation in the superconducting phase (see further discussions in Appendix~\ref{AppC}).

\begin{figure*}[!t]
    \centering
    \includegraphics[width=4 in]{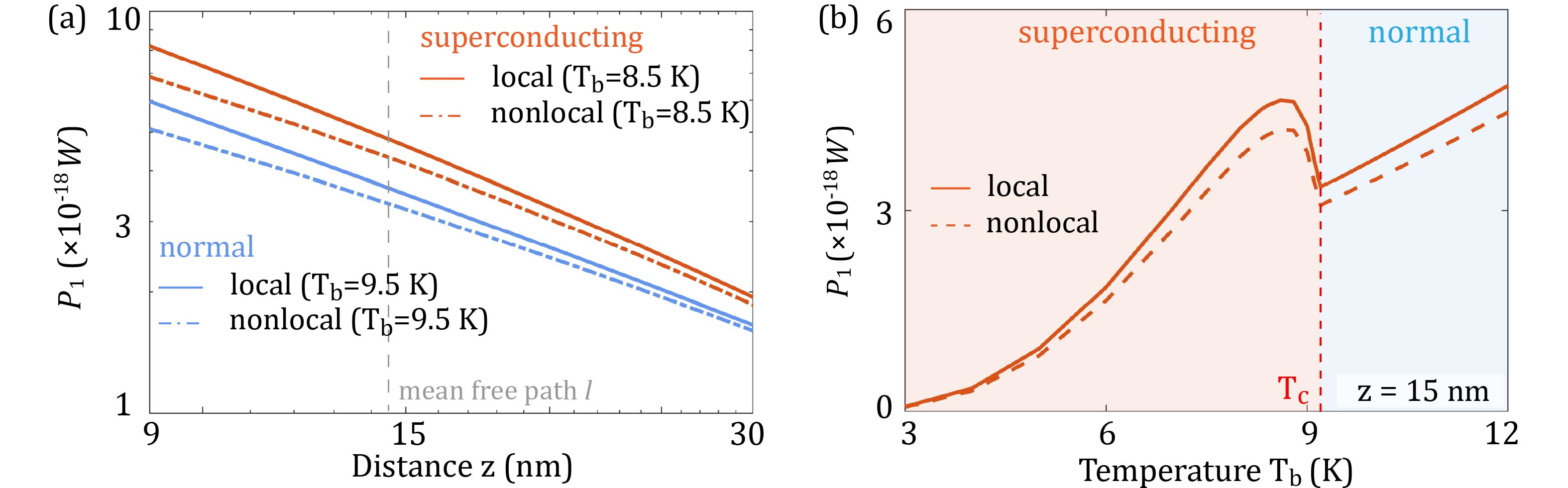}
    \caption{Superconducting coherence peak in NFRHT beyond the local response limit. (a) Dependence of NFRHT $P_1$ on the distance $z$ between the YIG nanoparticle and the niobium slab. Superconducting coherence peak in NFRHT becomes more prominent at smaller distance $z$ between the YIG nanoparticle and the niobium slab. (b) Nonlocal effects do not affect the superconducting coherence peak in NFRHT.}
    \label{fig:fig5}
\end{figure*}

The above mCDOS $\rho^m_c$ behaviors lead to interesting many-body radiative heat transfer effects $\Delta P$ at the superconducting phase transition. In Figs.~\ref{fig:fig4}(c,e), we consider a YIG nanoparticle array with lattice constant $D$ in the small $z=15\,\mathrm{nm}$ and large $z=150\,\mathrm{nm}$ regions. In general, NFRHT is enhanced by many-body effects, i.e., $\Delta P >0$. Meanwhile, as shown in Fig.~\ref{fig:fig4}(c), we find that the coherence of Bogoliubov quasiparticles enhances many-body effects $\Delta P$ right below the transition temperature in the small $z$ region. In contrast, as shown in Fig.~\ref{fig:fig4}(e), Cooper pair screening effects suppress the many-body effects $\Delta P$ in the superconducting phase. In addition, we note that, for NFRHT in extended nanoparticle arrays (large $D$) in the large $z$ region, Cooper pair screening effects can turn many-body effects from enhancement ($\Delta P > 0$) to suppression ($\Delta P < 0$). The above analysis isolates the different effects of Bogoliubov quasiparticles and Cooper pairs on many-body NFRHT.

\section{Discussions}
In summary, we reveal a superconducting coherence peak in NFRHT in the superconductor-ferromagnet system and isolate effects of Cooper pairs and Bogoliubov quasiparticles on NFRHT. Our predicted phenomenon can be experimentally observed in state-of-the-art experimental platforms, e.g., scanning thermal microscopy (SThM) probes~\cite{kim2015radiative,cui2017study}, which can probe nanoscale radiative heat transfer across a vacuum gap $\lesssim 10\,\mathrm{nm}$. Our results are important for developing nanoscale energy harvesting technologies, e.g., thermal diodes or rectifiers~\cite{kasali2020optimization,ben2013phase}, in superconducting circuits. 

Beyond this, our discussions of low-frequency magnetic LDOS and CDOS are also important for quantum dynamics of spin qubits near superconductors, e.g., spin dephasing~\cite{sun2025super} and relaxation~\cite{kelly2024superconductivity,li2024observation}.

To facilitate experimental observations, in Fig.~\ref{fig:fig5}, we briefly discuss the nonlocal effects and the distance dependence of the superconducting coherence peak in NFRHT. Our calculations are within the dipole approximation, and high-order multipolar effects are beyond the scope of the manuscript. For dirty BCS superconductors, we find that the nonlocality of the transverse permittivity $\varepsilon^\bot_{Nb}(k,\omega)$ described by the Mattis-Bardeen theory~\cite{mattis1958theory} becomes prominent at $k>1/l$, where $l$ is the mean free path of the superconductor (see Appendix~\ref{AppE}). In Fig.~\ref{fig:fig5}(a), we present the dependence of NFRHT $P_1$ on the distance $z$ between the YIG nanoparticle and the niobium slab. We study the distance dependence of $P_1$ when the niobium slab is in the superconducting phase at $T_b=8.5\,\mathrm{K}$ corresponding to the maximum of the superconducting coherence peak in Fig.~\ref{fig:fig3}(b), and when the niobium is in the normal phase at $T_b=9.5\,\mathrm{K}$. Here, we find that the superconducting coherence peak can become more prominent at smaller distance $z$. Furthermore, as shown in Fig.~\ref{fig:fig5}(b), our predicted superconducting coherence peak is not limited to the local response limit. We find that nonlocal effects do not affect the superconducting coherence peak and nonlocal effects only become appreciable at distance $z<l$ smaller than the mean free path. Calculation details for nonlocal NFRHT between the YIG nanoparticle and the superconducting niobium slab are provided in Appendix~\ref{AppE}.

\section*{Acknowledgments}
W.S. and Z.J. acknowledge support from the U.S. Department of Energy (DOE), Office of Basic Sciences under DE-SC0017717. Z.M.Z. acknowledges support from the U.S. Department of Energy (DOE), Office of Science, Basic Energy Sciences under Grant No. DE-SC0018369.

\appendix

\section{Many-Body Radiative Heat Transfer Theory}\label{AppA}
In this appendix, we provide detailed derivations of the many-body near-field radiative heat transfer (NFRHT) between ferromagnetic yttrium iron garnet (YIG) nanoparticles and the superconductor slab. Our derivations follow the many-body radiative heat transfer theory based on fluctuational electrodynamics~\cite{biehs2021near,messina2013fluctuation}.

We consider a system of $N$ YIG nanoparticles placed near a superconducting niobium slab. Superconductivity effects from the niobium slab are encoded in the near-field magnetic Green's function $\Bar{\Bar{G}}_m$. In the following derivation, we suppress the subscript $m$ to simplify the notation and only write $\Bar{\Bar{G}}$. We consider small nanoparticle sizes and employ the dipole approximation for YIG nanoparticles. The net power $P_i$ absorbed by the i\textit{th} nanoparticle is,
\begin{multline}\label{Pi_eq1}
    P_i=\langle \frac{d \Vec{m}_i}{dt} \cdot \Vec{H}_i(t) \rangle = 2 \int_0^\infty \frac{d\omega}{2\pi} \omega \int_0^\infty \frac{d\omega'}{2\pi} \\ \mathrm{Im} \Big[ \langle \Vec{m}_i(\omega) \cdot \Vec{H}_i^\dagger(\omega') \rangle e^{-i(\omega-\omega')t} \Big],
\end{multline}
where $\Vec{m}_i$ is the magnetic dipole moment of the i\textit{th} nanoparticle, $\Vec{H}_i^\dagger$ is the Hermitian conjugate of $\Vec{H}_i$, and $\Vec{H}_i$ is the magnetic field at the position of the i\textit{th} nanoparticle. $\Vec{m}_i$ and $\Vec{H}_i$ can be further decomposed into the fluctuating and induced components,
\begin{equation}
    \Vec{m}_i = \Vec{m}_i^{fl} + \Vec{m}_i^{ind}, \qquad \qquad
    \Vec{H}_i = \Vec{H}_i^{fl} + \Vec{H}_i^{ind},
\end{equation}
where $\Vec{m}_i^{ind}$ is the magnetic dipole moment induced by fluctuating fields $\Vec{H}_i^{fl}$, and $\Vec{H}_i^{ind}$ is the magnetic field induced by fluctuating dipole moment $\Vec{m}_i^{fl}$. We have,
\begin{align}
    &\begin{aligned}\label{Htot}
        \Vec{H}_i = \Vec{H}_i^{fl} + 4\pi k_0^2 \sum_j \Bar{\Bar{G}}_{ij} \Vec{m}_j,
    \end{aligned}\\
    &\begin{aligned}\label{mtot}
        \Vec{m}_i = \Vec{m}_i^{fl} + \Bar{\Bar{\alpha}}_i (\Vec{H}_i^{fl} + 4\pi k_0^2 \sum_{j \neq i} \Bar{\Bar{G}}_{ij} \Vec{m}_j),
    \end{aligned}
\end{align}
where $\Bar{\Bar{G}}_{ij}=\Bar{\Bar{G}}(\Vec{r}_i,\Vec{r}_j,\omega)$ and $\Bar{\Bar{\alpha}}_i$ is the polarizability tensor of the i\textit{th} YIG nanoparticle. From Eqs.~(\ref{Htot},\ref{mtot}), we can find that 
\begin{align}\label{m_matrix}
\begin{aligned}
        \begin{bmatrix}  \Vec{m}_1 \\ \vdots \\ \Vec{m}_N    \end{bmatrix} &= \Bar{\Bar{T}}^{-1} \begin{bmatrix}  \Vec{m}_1^{fl} \\ \vdots \\ \Vec{m}_N^{fl}    \end{bmatrix} + \Bar{\Bar{T}}^{-1} \Bar{\Bar{A}} \begin{bmatrix}  \Vec{H}_1^{fl} \\ \vdots \\ \Vec{H}_N^{fl}    \end{bmatrix} \\ &= \Bar{\Bar{M}} \begin{bmatrix}  \Vec{m}_1^{fl} \\ \vdots \\ \Vec{m}_N^{fl}    \end{bmatrix} + \Bar{\Bar{N}} \begin{bmatrix}  \Vec{H}_1^{fl} \\ \vdots \\ \Vec{H}_N^{fl}    \end{bmatrix},
\end{aligned}
\end{align}
\begin{align}\label{H_matrix}
\begin{aligned}
        \begin{bmatrix}  \Vec{H}_1 \\ \vdots \\ \Vec{H}_N    \end{bmatrix} &= \Bar{\Bar{D}} \Bar{\Bar{T}}^{-1} \begin{bmatrix}  \Vec{m}_1^{fl} \\ \vdots \\ \Vec{m}_N^{fl}    \end{bmatrix} + (\Bar{\Bar{I}}_{3N} + \Bar{\Bar{D}} \Bar{\Bar{T}}^{-1} \Bar{\Bar{A}}) \begin{bmatrix}  \Vec{H}_1^{fl} \\ \vdots \\ \Vec{H}_N^{fl}    \end{bmatrix}  \\ & = \Bar{\Bar{O}} \begin{bmatrix}  \Vec{m}_1^{fl} \\ \vdots \\ \Vec{m}_N^{fl}    \end{bmatrix} + \Bar{\Bar{P}} \begin{bmatrix}  \Vec{H}_1^{fl} \\ \vdots \\ \Vec{H}_N^{fl}    \end{bmatrix},
\end{aligned}
\end{align}
where $\Bar{\Bar{I}}_{3N}$ is the $3N \times 3N$ identity matrix. $\Bar{\Bar{T}}, \Bar{\Bar{A}}, \Bar{\Bar{D}}$ are $N \times N$ block matrices with each element being a $3\times 3$ matrix,
\begin{align}\label{matrix_element}
    \begin{aligned}
        \Bar{\Bar{T}}_{ij} = \delta_{ij} \Bar{\Bar{I}}_3 - 4\pi k_0^2 (1-\delta_{ij}) \Bar{\Bar{\alpha}}_i \Bar{\Bar{G}}_{ij},
    \end{aligned} \\
    \begin{aligned}
        \Bar{\Bar{A}}_{ij} = \delta_{ij} \Bar{\Bar{\alpha}}_i,
     \qquad
        \Bar{\Bar{D}}_{ij} = 4\pi k_0^2 \Bar{\Bar{G}}_{ij},
    \end{aligned}
\end{align}
where $\Bar{\Bar{I}}_3$ is the $3 \times 3$ identity matrix. Following Refs.~\cite{messina2013fluctuation}, we also define $\Bar{\Bar{B}}=\Bar{\Bar{D}} + \Bar{\Bar{A}}^{-1} \Bar{\Bar{T}}$. $\Bar{\Bar{B}}$ is also a $N \times N$ block matrix with element
\begin{equation}\label{matrix_element2}
    \Bar{\Bar{B}}_{ij} = \delta_{ij} [\Bar{\Bar{\alpha}}_i^{-1} + \Bar{\Bar{D}}_{ij}].
\end{equation}
Substituting Eqs.~(\ref{m_matrix},\ref{H_matrix}) into Eq.~(\ref{Pi_eq1}), we find,
\begin{align}\label{Pi_eq2}
\begin{aligned}
    &\langle \Vec{m}_i(\omega) \cdot \Vec{H}_i^\dagger(\omega') \rangle = \sum_\alpha \sum_{\beta\beta'} \sum_{jj'} \Big( [\Bar{\Bar{M}}_{ij}]_{\alpha\beta} \langle [\Vec{m}_j^{fl}]_\beta [\Vec{m}_{j'}^{fl\dagger}]_{\beta'} \rangle \\ & \qquad \qquad  [\Bar{\Bar{O}}_{j'i}^\dagger]_{\beta'\alpha} + [\Bar{\Bar{N}}_{ij}]_{\alpha\beta} \langle [\Vec{H}_j^{fl}]_\beta [\Vec{H}_{j'}^{fl\dagger}]_{\beta'} \rangle [\Bar{\Bar{P}}_{j'i}^\dagger]_{\beta'\alpha} \Big),
\end{aligned}
\end{align}
where $[\Bar{\Bar{M}}_{ij}]_{\alpha\beta}$ denotes the element of the $3 \times 3$ matrix $\Bar{\Bar{M}}_{ij}$ and $[\Vec{m}_j^{fl}]_\beta$ denotes the element of the $3 \times 1$ vector $\Vec{m}_j^{fl}$. In the above derivation, we use $\langle [\Vec{m}_j^{fl}]_\beta [\Vec{H}_{j'}^{fl\dagger}]_{\beta'} \rangle = 0$, i.e., there is no correlation between the magnetic dipole fluctuations and magnetic filed fluctuations, to simplify the expression. The fluctuations $\langle [\Vec{m}_j^{fl}]_\beta [\Vec{m}_{j'}^{fl\dagger}]_{\beta'} \rangle$ and $\langle [\Vec{H}_j^{fl}]_\beta [\Vec{H}_{j'}^{fl\dagger}]_{\beta'} \rangle$ can be evaluated using the fluctuation-dissipation theorem,
\begin{align}
    &\begin{aligned}\label{fdt_dipole}
        \langle [\Vec{m}_j^{fl}]_\beta [\Vec{m}_{j'}^{fl\dagger}]_{\beta'} \rangle = & 4\pi\hbar \delta(\omega-\omega') \\ & \quad \delta_{jj'} (n_j + \frac{1}{2}) \bigg( \frac{\Bar{\Bar{\alpha}}_{j}-\Bar{\Bar{\alpha}}_{j}^\dagger}{2i}\bigg)_{\beta \beta'},
    \end{aligned}\\
    &\begin{aligned}\label{fdt_field}
        \langle [\Vec{H}_j^{fl}]_\beta [\Vec{H}_{j'}^{fl\dagger}]_{\beta'} \rangle & =  4\pi k_0^2 4\pi\hbar \delta(\omega-\omega') \\ & \quad (n_b + \frac{1}{2}) \bigg( \frac{[\Bar{\Bar{G}}]_{jj'}-[\Bar{\Bar{G}}]_{j'j}^\dagger}{2i} \bigg)_{\beta \beta'},
    \end{aligned}
\end{align}
where $n_i$ and $n_b$ are the mean photon number at the i\textit{th} particle temperature $T_i$ and environment temperature $T_b$,
\begin{equation}
    n_i=\frac{1}{e^{\hbar\omega/k_B T_i}-1}, \qquad \qquad n_b=\frac{1}{e^{\hbar\omega/k_B T_b}-1}.
\end{equation}
Substituting Eqs.~(\ref{fdt_dipole},\ref{fdt_field}) into Eq.~(\ref{Pi_eq2}), we obtain,
\begin{align}\label{mb_th_2}
\begin{aligned}
    & \langle \Vec{m}_i(\omega) \cdot \Vec{H}_i^\dagger(\omega') \rangle = 4\pi\hbar\delta(\omega-\omega') \Big\{ \sum_j (n_j+\frac{1}{2}) \\ & \mathrm{Tr} \big[ \Bar{\Bar{M}}_{ij} \Bar{\Bar{\chi}}_{jj} \Bar{\Bar{O}}_{ji}^\dagger \big] + 4\pi k_0^2 (n_b+\frac{1}{2}) \sum_{jj'} \mathrm{Tr} \big[ \Bar{\Bar{N}}_{ij} \mathrm{Im} [\Bar{\Bar{G}}_{jj'}] \Bar{\Bar{P}}_{j'i}^\dagger \big] \Big\},
\end{aligned}
\end{align}
where $\Bar{\Bar{\chi}}_{jj} = (\Bar{\Bar{\alpha}}_{j}-\Bar{\Bar{\alpha}}_{j}^\dagger)/2i$. Since the photonic environment near BCS superconductor is reciprocal, we also have $([\Bar{\Bar{G}}]_{jj'}-[\Bar{\Bar{G}}]_{j'j}^\dagger)/2i = \mathrm{Im} [\Bar{\Bar{G}}_{jj'}]$.

In the main text, we consider all particles to have the same temperature. Substitute Eqs.~(\ref{matrix_element}-\ref{matrix_element2}) into Eq.~(\ref{mb_th_2}) and simplify Eq.~(\ref{mb_th_2}) by using the second law of thermodynamics, we find,
\begin{multline}\label{P_many_body}
    P_i=8 \int_0^\infty d\omega \hbar \omega k_0^2  (n_b-n_i) \, \mathrm{Im} \big\{ \sum_{jj'}^N \mathrm{Tr} \big[ [\Bar{\Bar{T}}^{-1}]_{ij} [\Bar{\Bar{A}}]_{jj} \\ \mathrm{Im} [\Bar{\Bar{G}}]_{jj'} [\Bar{\Bar{A}}^{\dagger}]_{j'j'} [\Bar{\Bar{T}}^{-1,\dagger}]_{j'i} [\Bar{\Bar{B}}^{\dagger}]_{ii} \big] \big\},
\end{multline}
which is Eq.~(\ref{many-body}) in the main text. For the special case $N=1$, we find Eq.~(\ref{P_many_body}) is reduced to,
\begin{equation}\label{P_single}
    P_1=8  \int_0^\infty d\omega \hbar \omega k_0^2 (n_b-n_1) \, \mathrm{Tr} [\mathrm{Im}[\Bar{\Bar{G}}_{11}] \mathrm{Im}[\Bar{\Bar{\alpha}}_1]],
\end{equation}
which is the net power absorbed by a single nanoparticle from the environment (Eq.~(\ref{P1}) in the main text).

\section{Calculations of dyadic Green's functions}~\label{AppB}
In this section, we provide the expressions for the dyadic Green's functions. The dyadic Green's functions can be separated into the free space part and the reflected part $\Bar{\Bar{G}}=\Bar{\Bar{G}}^0+\Bar{\Bar{G}}^r$. The free space part is~\cite{novotny2012principles},
\begin{widetext}
\begin{equation}\label{DGreen0}
    \Bar{\Bar{G}}_m^0 (\Vec{r}_i,\Vec{r}_j,\omega) = \Bar{\Bar{G}}_e^0 (\Vec{r}_i,\Vec{r}_j,\omega) = \frac{e^{ik_0D}}{4\pi D}\Big[\big(1+\frac{ik_0D-1}{k_0^2D^2}\big)\Bar{\Bar{I}}_3 + \frac{3-3ik_0D-k_0^2D^2}{k_0^2D^2} \frac{\Vec{D}\Vec{D}}{D^2}\Big],
\end{equation}
where $\Vec{D}=\Vec{r}_i - \Vec{r}_j$, and the subscript $m,e$ denotes the magnetic and electric dyadic Green's functions. In the near-field of normal conductors or superconductors, low-frequency dyadic Green's functions are generally dominated by the reflected part $\Bar{\Bar{G}}^r$~\cite{volokitin2007near}, i.e., $\Bar{\Bar{G}}\approx \Bar{\Bar{G}}^r$. The reflected part of dyadic Green's functions is~\cite{novotny2012principles},
\begin{align}
    \Bar{\Bar{G}}_m^r (\Vec{r}_i,\Vec{r}_j,\omega)&=\frac{i}{8 \pi^2}\int \frac{d \Vec{q}}{k_z} e^{i \Vec{q}\cdot (\Vec{r}_i-\Vec{r}_j)} e^{i k_z(z_i+z_j)} \biggl( \frac{r_p}{q^2}\begin{bmatrix} q_y^2&-q_x q_y&0\\-q_x q_y&q_x^2&0\\0&0&0 \end{bmatrix}+ \frac{r_s}{k_0^2q^2}\begin{bmatrix} -q_x^2 k_z^2 & -q_x q_y k_z^2 & -q_x k_z q^2\\-q_x q_y k_z^2 & -q_y^2 k_z^2 & -q_y k_z q^2\\q^2 q_x k_z & q^2 q_y k_z & q^4 \end{bmatrix} \biggr),\label{DGreenm}\\
    \Bar{\Bar{G}}_e^r (\Vec{r}_i,\Vec{r}_j,\omega)&=\frac{i}{8 \pi^2}\int \frac{d \Vec{q}}{k_z} e^{i \Vec{q} \cdot (\Vec{r}_i-\Vec{r}_j)} e^{i k_z(z_i+z_j)} \biggl( \frac{r_s}{q^2}\begin{bmatrix} q_y^2&-q_x q_y&0\\-q_x q_y&q_x^2&0\\0&0&0 \end{bmatrix}+ \frac{r_p}{k_0^2q^2}\begin{bmatrix} -q_x^2 k_z^2 & -q_x q_y k_z^2 & -q_x k_z q^2\\-q_x q_y k_z^2 & -q_y^2 k_z^2 & -q_y k_z q^2\\q^2 q_x k_z & q^2 q_y k_z & q^4 \end{bmatrix} \biggr),\label{DGreene}
\end{align}
\end{widetext}
where $\Vec{q}$ is the in-plane wavevector, $k_z=\sqrt{k_0^2-q^2}$, $z_i,z_j$ are the distance between $\Vec{r}_i,\Vec{r}_j$ and the niobium slab interface, and $r_s, r_p$ are the Fresnel reflection coefficients. In the main text, we use Eqs.~(\ref{DGreen0}-\ref{DGreene}) to calculate the dyadic Green's functions near the superconducting niobium slab.

In the main text, we consider the local permittivity $\varepsilon_{Nb}(\omega)$ of the superconducting niobium. In this case, the Fresnel reflection coefficients are,
\begin{equation}
    r_s(q)=\frac{k_z-k_z'}{k_z+k_z'}, \qquad r_p(q)=\frac{\varepsilon_{\mathrm{Nb}}k_z-k_z'}{\varepsilon_{\mathrm{Nb}}k_z+k_z'},
\end{equation}
where $k_z'=\sqrt{\varepsilon_{\mathrm{Nb}}k_0^2-q^2}$ and $\varepsilon_{Nb}$ is the local permittivity of superconducting niobium described by the local response limit of the Mattis-Bardeen theory (see Appendix~\ref{AppC}).

\subsection{Near-Field Approximation}
Meanwhile, we can also find approximate expressions of the dyadic Green's functions to explicate the physics. In the near-field $z\ll 1/k_0$, magnetic dyadic Green's function is commonly dominated by s-polarized evanescent waves with momentum $q \gg k_0$~\cite{joulain2003definition,chapuis2008effects,zhang2007nano,volokitin2007near}. This can be understood by considering that the factor $e^{2ik_z z}=e^{-2z\sqrt{q^2-k_0^2}}$ in Eq.~(\ref{DGreenm}) acts as a momentum cutoff function for the integral. This cutoff function makes the integral in Eq.~(\ref{DGreenm}) to be dominated by contributions from $q \sim q_c \approx 1/(2z) \gg k_0$. In Eq.~(\ref{DGreenm}), $r_p$ term is proportional to $(q/k_0 )^0$, while the $r_s$ term is proportional to $(q/k_0 )^2$. This reveals that near-field magnetic dyadic Green's functions are dominated by the s-polarized evanescent waves. Therefore, from Eq.~(\ref{DGreenm}), in the near-field regime $z\ll 1/k_0$, we have 
\begin{equation}\label{dyadG_App}
    \Bar{\Bar{G}}_m (\Vec{r}_1,\Vec{r}_1,\omega) \approx \frac{i}{8 \pi k_0^2}\int \frac{q dq}{k_z} e^{-2q z} r_s(q) \begin{bmatrix} -k_z^2 & 0 & 0\\0 & -k_z^2 & 0\\0 & 0 & 2q^2 \end{bmatrix},
\end{equation}
where we consider $z_1=z$. This is Eq.~(\ref{dyadG}) used in the main text for explicating the physics underlying the superconducting coherence peak. 

From Eq.~(\ref{dyadG_App}), the mLDOS $\rho_l^m$ is,
\begin{equation}\label{mldos_approx}
    \rho^m_l \approx \frac{1}{2\pi^2 \omega} \int_0^\infty  dq \, e^{-2q z} q^2 \mathrm{Im}[r_s(q)],
\end{equation}
where we considered $k_z\approx iq$. From Eq.~(\ref{mldos_approx}), we can find $\partial \rho^m_l/\partial q$ in Eqs.~(\ref{smallz},~\ref{largez}).

Similarly, at $z\ll1/k_0$, we can also find the approximate expressions for magnetic CDOS $\rho^m_c$. Without loss of generality, we can consider $\rho^m_c$ at two points aligned along the x-axis. We have, 
\begin{equation}\label{mcdos}
    \rho^m_c \approx \frac{1}{4\pi^3 \omega} \int_0^\infty  dq \int_0^{2\pi} d\theta \, e^{q (iD \cos{\theta}-2z)} q^2 \mathrm{Im}[r_s(q)],
\end{equation}
where $D$ is the distance between the two points. 

Similar to mLDOS, mCDOS Eq.~(\ref{mcdos}) is also dominated by contributions from evanescent waves with momentum $q \sim q_c \sim 1/2z \gg 1/k_0$. In this regime, we can then find approximate expressions of $r_s$. For the niobium slab in the normal phase or in the small $z$ region, we have $q_c/k_0\sim (2zk_0)^{-1} \gg |\varepsilon_{Nb}| \gg 1$. Therefore, we have 
\begin{equation}\label{mcdos_smallz}
    \rho^m_c \approx \frac{\omega \mathrm{Im} [\varepsilon_\mathrm{Nb}-1]}{16\pi^3c^2} \int_0^\infty dq \int_0^{2\pi} d\theta \, e^{q(iD\cos{\theta}-2z)}.
\end{equation}
At distance $D>2z$, we can find $\rho^m_c \sim D^{-1}$. 

Meanwhile, in the superconducting phase, Cooper pairs contribute to very strong screening effects (i.e., large $-\mathrm{Re}[\varepsilon_{Nb}]$) in the superconductor EM response. Therefore, for the niobium slab in the superconducting phase or in the large $z$ region, we have $-\mathrm{Re}[\varepsilon_{Nb}]  \gtrsim q_c/k_0\sim (2zk_0)^{-1} \gg 1$. In this case, 
\begin{equation}\label{mcdos_largez}
    \rho^m_c \approx \frac{c}{2\pi^3\omega^2} \mathrm{Im} [\frac{i}{\sqrt{\varepsilon_{Nb}}}] \int_0^\infty dq \int_0^{2\pi} d\theta \, q^3 e^{q(iD\cos{\theta}-2z)}.
\end{equation}
At distance $D>2z$, we can find $\rho^m_c \sim -D^{-5}$. 

Comparing Eqs.~(\ref{mcdos_smallz},~\ref{mcdos_largez}), we explain that Cooper pair screening effects suppress the spatial coherence of near-field thermal fluctuations, which is demonstrated in Figs.~\ref{fig:fig4}(b,d) in the main text.

\section{Electromagnetic Response of BCS Superconductors}\label{AppC}
In this Appendix, we provide the model and parameters for the permittivity $\varepsilon_{\mathrm{Nb}}$ of the niobium slab considered in the main text. Niobium typically has the superconducting phase transition temperature at $T_c=9.2\,\mathrm{K}$. In the normal phase, we consider the Drude model for $\varepsilon_{\mathrm{Nb}}(\omega)$,
\begin{equation}\label{drude}
    \varepsilon_{\mathrm{Nb}}(\omega) = 1-\frac{\omega_p^2\tau^2}{\omega^2\tau^2+1} + i \frac{\sigma_n}{\varepsilon_0\omega(1+\omega^2\tau^2)},
\end{equation}
where we consider the normal state conductivity $\sigma_n = 2\times 10^7 \,\mathrm{S/m}$, plasma frequency $\omega_p=8.8\times 10^{15} \,\mathrm{Hz}$~\cite{pronin1998direct,gubin2005dependence}, and corresponding electron relaxation time $\tau = \sigma_n/(\varepsilon_0\omega_p^2) \approx 2.9 \times 10^{-14} \,\mathrm{s}$. 

\subsection{Local response limit of Mattis-Bardeen theory}
In the superconducting phase, we consider the Mattis-Bardeen theory for niobium permittivity $\varepsilon_{\mathrm{Nb}}$~\cite{mattis1958theory}. Mattis-Bardeen theory is applicable to weak-coupling s-wave BCS superconductors with arbitrary purity and with local or nonlocal EM response, and matches well with experimental measurements~\cite{klein1994conductivity}.

For conventional dirty BCS superconductors, nonlocal effects are not prominent at $k < l^{-1}$ (see discussions in Appendix~\ref{AppE}), i.e., $\varepsilon^\bot_{\mathrm{Nb}}(k<l^{-1},\omega) \approx \varepsilon_{\mathrm{Nb}}(0,\omega)$, where $l$ is the mean free path and $\varepsilon^\bot_{\mathrm{Nb}}(k,\omega)$ is the transverse component of nonlocal permittivity given by the Mattis-Bardeen theory~\cite{mattis1958theory}. As discussed in Appendix~\ref{AppB}, NFRHT in our configurations is determined by evanescent waves with momentum $\sim q_c \approx 1/(2z) <1/l$, indicating that nonlocal effects are not important for our configurations in the main text. Therefore, in the main text, we consider the local response limit $\varepsilon_{\mathrm{Nb}}(\omega)= \varepsilon^\bot_{\mathrm{Nb}}(k\rightarrow 0,\omega)$ in our calculations. We extend our results to account for nonlocal effects in EM response in Appendix~\ref{AppE}. The permittivity $\varepsilon_{\mathrm{Nb}}(\omega)$ in the local response limit from the Mattis-Bardeen theory is,
\begin{widetext}
\begin{subequations}\label{eps_nb_local}
    \begin{equation}
    \varepsilon_{\mathrm{Nb}}(\omega) = 1-\frac{\sigma_n K_1(\omega)}{\hbar \varepsilon_0 \omega^2} + i \frac{\sigma_n K_2(\omega)}{\hbar \varepsilon_0 \omega^2},
    \end{equation} 
    \begin{multline}
    K_1(\omega) =  \int_{\Delta}^{\infty} dE \big[f(E)-f(E+\hbar\omega)\big] \big[g_1(E)+1\big] \frac{a_-}{1+a^2_-} 
    - \int_{\Delta}^{\infty} dE \big[1-f(E)- f(E+\hbar\omega)\big] \big[g_1(E)-1\big] \frac{a_+}{1+a^2_+}  \\   
    + \int_{max\{\Delta-\hbar\omega, -\Delta\}}^{\Delta} dE \big[1-2f(E+\hbar\omega)\big] \Big\{ g_2(E) \frac{|a_1| + 1}{a_2^2+(|a_1| + 1)^2} + \frac{a_2}{a_2^2+(|a_1| + 1)^2} \Big\} \\
    + \frac{\Theta(\hbar\omega-2\Delta)}{2} \int_{\Delta-\hbar\omega}^{-\Delta} dE \big[1-2f(E+\hbar\omega)\big] \Big\{ \big[g_1(E)+1\big] \frac{a_-}{1+a^2_-} - \big[g_1(E)-1\big] \frac{a_+}{1+a^2_+} \Big\} ,
    \end{multline} 
    \begin{multline}
    K_2(\omega) = \int_{\Delta}^{\infty} dE \big[f(E) - f(E+\hbar\omega) \big] \Big\{ \big[g_1(E)+1\big] \frac{1}{1+a_-^2} + \big[g_1(E)-1\big]\frac{1}{1+a_+^2} \Big\} \\
    - \frac{\Theta(\hbar\omega-2\Delta)}{2} \int_{\Delta-\hbar\omega}^{-\Delta} dE \big[1-2f(E+\hbar\omega)\big] \Big\{ \big[g_1(E)+1\big] \frac{1}{1+a_-^2} + \big[g_1(E)-1\big]\frac{1}{1+a_+^2} 
    \Bigg\},
    \end{multline}
\end{subequations}
where
\begin{subequations}\label{eps_nb_local_2}
    \begin{equation}
        f(E) = 1/(1+e^{E/k_bT}), 
    \end{equation}
    \begin{equation}
        g_1 = \frac{E^2+\Delta^2+\hbar\omega E}{\sqrt{E^2-\Delta^2}\sqrt{(E+\hbar\omega)^2-\Delta^2}}, \qquad g_2 = \frac{E^2+\Delta^2+\hbar\omega E}{\sqrt{\Delta^2 - E^2}\sqrt{(E+\hbar\omega)^2-\Delta^2}},
    \end{equation}
    \begin{equation}
        a_1 = \frac{\sqrt{E^2-\Delta^2}\, \tau}{\hbar }, \qquad a_2 = \frac{\sqrt{(E+\hbar\omega)^2-\Delta^2} \, \tau}{\hbar }, \qquad  a_-=a_2-a_1, \qquad a_+=a_1+a_2,
    \end{equation}
\end{subequations}
\end{widetext}
where $\Delta$ is the superconducting bandgap, $\Theta$ is the Heaviside step function, and $k_b$ is the Boltzmann constant. We note that Eqs.~(\ref{eps_nb_local},~\ref{eps_nb_local_2}) are identical to expressions in Ref.~\cite{zimmermann1991optical}. For $\tau \approx 2.9 \times 10^{-14} \,\mathrm{s}$ and Fermi velocity $v_F\approx 5\times 10^5 \,\mathrm{m/s}$~\cite{mattheiss1970electronic}, we find that the mean free path $l=v_F\tau\approx 14.5\,\mathrm{nm}$ is smaller than the BCS coherence length $l < \xi_0$, corresponding to a moderately dirty niobium. In Eq.~(\ref{eps_nb_local}), at low frequencies $\omega \ll \omega_g$, the screening effects $\mathrm{Re} \, \varepsilon_{\mathrm{Nb}}$ are dominated by the Cooper pair response, while dissipation $\mathrm{Im} \, \varepsilon_{\mathrm{Nb}}$ is dominated by the Bogoliubov quasiparticle response~\cite{dressel2013electrodynamics}. We note that the superconducting bandgap $\Delta$ affects the permittivity $\varepsilon_{\mathrm{Nb}}$, which then determines the NFRHT $P_1$. In the limit of $\Delta \rightarrow 0$, Eq.~(\ref{eps_nb_local}) reduces to the Drude model.

\section{Polarizability Tensor of YIG and Gold Nanoparticles}
In this Appendix, we provide the YIG permeability tensor $\Bar{\Bar{\mu}}_{\mathrm{YIG}}$ described by the Landau-Liftshitz-Gilbert equations~\cite{cullity2011introduction}. We start from the free energy $H$ of the spherical ferromagnetic YIG nanoparticle that contains the Zeeman energy and anisotropy energy,
\begin{equation}\label{free_e}
    H = - \mu_0 M_s \Vec{\mathcal{M}} \cdot \Vec{\mathbf{H}}_{\mathrm{ext}} - K_{x} m_x^2,
\end{equation}
where $\Vec{\mathcal{M}}$ is the normalized magnetization direction vector, $\mathcal{M}_x$ is the $x$ component of $\Vec{\mathcal{M}}$, $M_s$ is the saturation magnetization, and $K_x$ is the anisotropy constant. In Eq.~(\ref{free_e}), we consider the YIG nanoparticle to have uniaxial anisotropy with the anisotropy axis along the x-direction. From Eq.~(\ref{free_e}), we find the effective field $\Vec{\mathbf{H}}_{\mathrm{eff}}$ is,
\begin{multline}\label{H_eff}
    \Vec{\mathbf{H}}_{\mathrm{eff}}= -\frac{1}{\mu_0 M_s} \frac{\partial H}{\partial \Vec{\mathcal{M}}} =  \Vec{\mathbf{H}}_{\mathrm{ext}} + 2\frac{K_{\parallel}}{\mu_0 M_s}m_x\mathbf{\hat{x}} \\ =\Vec{\mathbf{H}}_{\mathrm{ext}} + H_k m_x\mathbf{\hat{x}}.
\end{multline}
In this work, YIG nanoparticles are not subject to the external magnetic field, i.e., $\Vec{\mathbf{H}}_{\mathrm{ext}}=0$. Therefore, the equilibrium magnetization is along the $\hat{\mathbf{x}}$ direction. Substitute Eq.~(\ref{H_eff}) in to the Landau-Liftshitz-Gilbert equation, we have,
\begin{equation}\label{LLG}
    \frac{d\Vec{\mathcal{M}}}{dt}=-\mu_0\gamma \Vec{\mathcal{M}} \times \mathbf{H}_{\mathrm{eff}}+\alpha_{loss} \Vec{\mathcal{M}}\times\frac{d\Vec{\mathcal{M}}}{dt},
\end{equation}
where $\alpha_{loss}$ is the Gilbert damping constant describing the loss of the ferromagnet and $\gamma$ is the gyromagnetic ratio. The permeability tensor is then conventionally solved from Eq.~(\ref{LLG}) by considering the perturbation of $\delta \Vec{\mathcal{M}}$ under a small external magnetic field $\delta \Vec{\mathbf{H}}$~\cite{cullity2011introduction}. We find the magnetic resonance frequency (in the absence of external magnetic fields) $\omega_r=\gamma\mu_0 H_k$, and the permeability tensor,
\begin{equation}
    \Bar{\Bar{\mu}}_{\mathrm{YIG}}= 
    \begin{bmatrix}
         1 & 0 & 0 \\
         0 & 1+\frac{\omega_m (\omega_r-i\alpha_{loss}\omega)}{(\omega_r-i\alpha_{loss}\omega)^2-\omega^2} & -i\frac{\omega_m \omega}{(\omega_r-i\alpha_{loss}\omega)^2-\omega^2} \\
         0 & i\frac{\omega_m \omega}{(\omega_r-i\alpha_{loss}\omega)^2-\omega^2} & 1+\frac{\omega_m (\omega_r-i\alpha_{loss}\omega)}{(\omega_r-i\alpha_{loss}\omega)^2-\omega^2} \\
    \end{bmatrix},
\end{equation}
where $\omega_m=\gamma\mu_0 M_s$. In the main text, we consider $\alpha_{loss}=0.01$, $4\pi M_s=1819\,\mathrm{G}$, and $H_k=600\,\mathrm{G}$ comparable to experimental values~\cite{sanchez2002particle,rajendran2006size}. The magnetic polarizability tensor for the gyromagnetic YIG nanoparticle can be obtained from $\Bar{\Bar{\mu}}_{\mathrm{YIG}}$ using the method in Ref.~\cite{sihvola1994dielectric},
\begin{multline}
    \Bar{\Bar{\alpha}}_{\mathrm{YIG}}= r^3 \\
    \begin{bmatrix}
         \frac{-1+\mu_{xx}}{2+\mu_{xx}}         & 0 & 0 \\
         0 & \frac{(\mu_{yy}-1)(\mu_{zz}+2)+u_{yz}^2}{(\mu_{yy}+2)(\mu_{zz}+2)+u_{yz}^2} & \frac{3\mu_{yz}}{(\mu_{yy}+2)(\mu_{zz}+2)+u_{yz}^2} \\
         0 & -\frac{3\mu_{yz}}{(\mu_{yy}+2)(\mu_{zz}+2)+u_{yz}^2} & \frac{(\mu_{zz}-1)(\mu_{yy}+2)+u_{yz}^2}{(\mu_{yy}+2)(\mu_{zz}+2)+u_{yz}^2} \\
    \end{bmatrix}.
\end{multline}

\begin{figure}[!t]
    \centering
    \includegraphics[width=3.4 in]{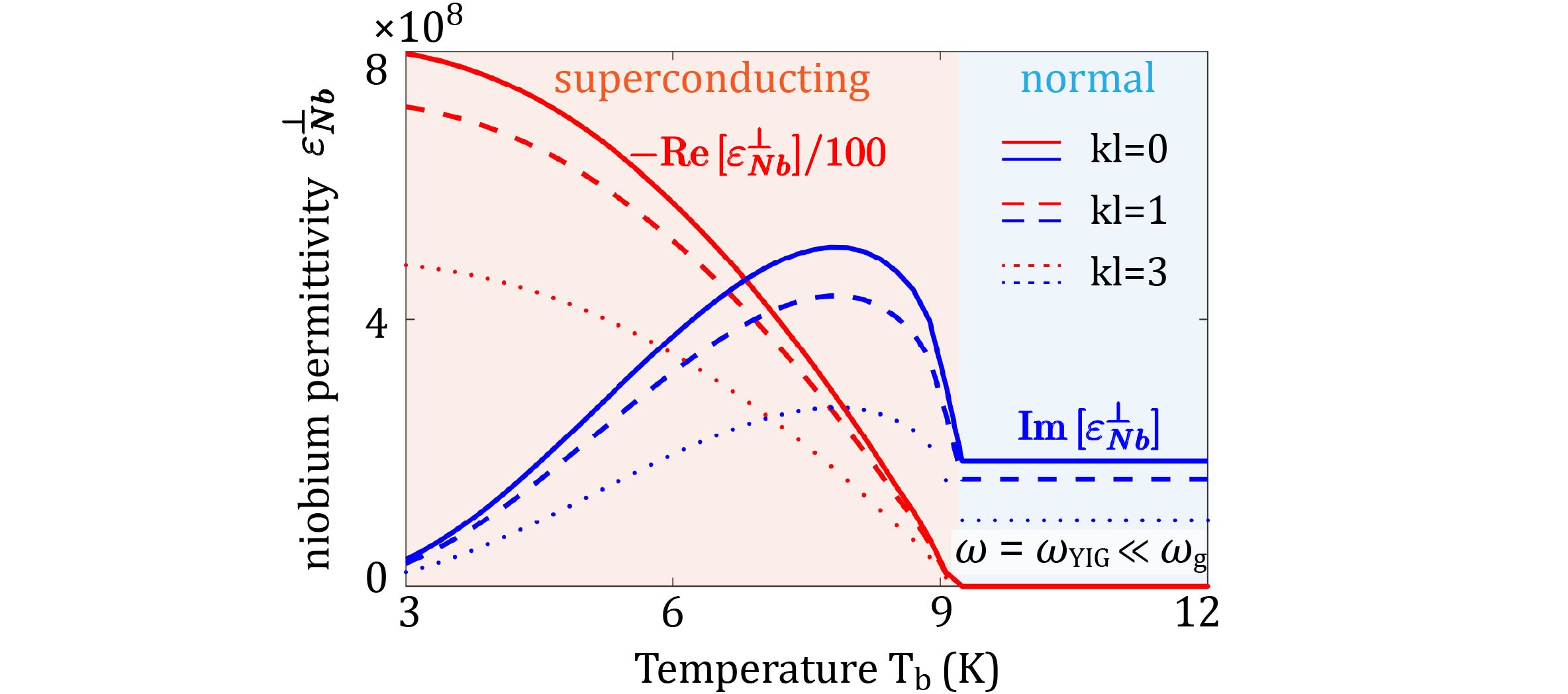}
    \caption{Nonlocal transverse permittivity $\varepsilon^\bot_{Nb}(k,\omega)$ predicted by the Mattis-Bardeen theory Eq.~(\ref{eps_nb}).}
    \label{fig:fig6}
\end{figure}

\subsection{Polarizability tensor for gold nanoparticles}
In the main text, we also consider NFRHT for gold nanoparticles to compare with ferromagnetic YIG nanoparticles. In this work, we consider the Drude model for gold permittivity $\varepsilon_{\mathrm{au}}$. We consider the gold conductivity $\sigma_{\mathrm{au}}=5\times 10^7\,\mathrm{S/m}$ and the plasma frequency $\omega_p\approx 1.3 \times 10^{16}\,\mathrm{Hz}$.

In our calculations, we consider contributions from both magnetic and electric dipole moment fluctuations of the gold nanoparticle to NFRHT. We note that the NFRHT between the gold nanoparticle and the superconducting niobium slab is generally dominated by the magnetic fluctuations. The electric and magnetic polarizability of a gold nanoparticle with radius $r$ is~\cite{khosravi2024giant},
\begin{equation}
    \alpha_{\mathrm{au},e} = r^3 \frac{\varepsilon_{\mathrm{au}}-1}{\varepsilon_{\mathrm{au}}+2}, \quad \alpha_{\mathrm{au},m} = \frac{r^5}{30} \big( \frac{\omega}{c} \big)^2 (\varepsilon_{\mathrm{au}}-1).
\end{equation}
Meanwhile, we obtain the electric dipole moment contributions to radiative heat transfer by changing the magnetic polarizability tensor and magnetic Green's function in Eqs.~(\ref{P_many_body},\ref{P_single}) to their electric counterparts.

\section{Nonlocal Effects on Radiative Heat Transfer Near Superconductors}\label{AppE}
In this appendix, we extend our discussions beyond the local response limit. Here, we provide the transverse nonlocal permittivity $\varepsilon^\bot_{\mathrm{Nb}}(k,\omega)$ of superconducting niobium described by the Mattis-Bardeen theory~\cite{mattis1958theory,popel1989surface} and provide details of the calculations of nonlocal NFRHT in Fig.~\ref{fig:fig5}.

As discussed in Appendix~\ref{AppB}, near-field magnetic dyadic Green's function is dominated by contributions from s-polarized evanescent waves~\cite{joulain2003definition,chapuis2008effects}. Therefore, for calculating nonlocal NFRHT in Fig.~\ref{fig:fig5}, we only consider contributions from the $r_s$ term in Eq.~(\ref{DGreenm}). For materials with the transverse nonlocal permittivity $\varepsilon^\bot(k,\omega)$, the Fresnel reflection coefficient $r_s(q)$ is~\cite{ford1984electromagnetic},
\begin{equation}
    r_s(q,\omega) = \frac{2ik_z\int_0^\infty \frac{d\kappa}{\varepsilon^\bot_{Nb}(\sqrt{q^2+\kappa^2},\omega)-q^2/k_0^2-\kappa^2/k_0^2} - \pi k_0^2}{2ik_z\int_0^\infty \frac{d\kappa}{\varepsilon^\bot_{Nb}(\sqrt{q^2+\kappa^2},\omega)-q^2/k_0^2-\kappa^2/k_0^2}  + \pi k_0^2}.
\end{equation}

For niobium in the normal phase, we consider the transverse nonlocal permittivity described by the Lindhard formula~\cite{ford1984electromagnetic,sun2023limits}. Meanwhile, for niobium in the superconducting phase, the general form of the transverse component of nonlocal permittivity $\varepsilon^\bot_{\mathrm{Nb}}(k,\omega)$ is given by the Mattis-Bardeen theory~\cite{mattis1958theory,popel1989surface},
\begin{widetext}
\begin{subequations}\label{eps_nb}
    \begin{equation}
    \varepsilon^\bot_{\mathrm{Nb}}(k,\omega) = 1-\frac{3\sigma_n K_1(k,\omega)}{\hbar \varepsilon_0 \omega^2} + i \frac{3\sigma_n K_2(k,\omega)}{\hbar \varepsilon_0 \omega^2},
    \end{equation} 
    \begin{multline}
    K_1(k,\omega) =  \beta \Bigg\{ \int_{\Delta}^{\infty} dE \big[f(E)-f(E+\hbar\omega)\big] \big[g_1(E)+1\big] S(a_-,\beta) \\
    - \int_{\Delta}^{\infty} dE \big[1-f(E)- f(E+\hbar\omega)\big] \big[g_1(E)-1\big] S(a_+,\beta) \\   
    + \int_{max\{\Delta-\hbar\omega, -\Delta\}}^{\Delta} dE \big[1-2f(E+\hbar\omega)\big] \Big\{ g_2(E) R(a_2,|a_1| + \beta) + S(a_2,|a_1| + \beta) \Big\} \\
    + \frac{\Theta(\hbar\omega-2\Delta)}{2} \int_{\Delta-\hbar\omega}^{-\Delta} dE \big[1-2f(E+\hbar\omega)\big] \Big\{ \big[g_1(E)+1\big] S(a_-,\beta) - \big[g_1(E)-1\big]S(a_+,\beta) \Big\} 
    \Bigg\},
    \end{multline} 
    \begin{multline}
    K_2(k,\omega) =  \beta \Bigg\{ \int_{\Delta}^{\infty} dE \big[f(E) - f(E+\hbar\omega) \big] \Big\{ \big[g_1(E)+1\big] R(a_-,\beta) + \big[g_1(E)-1\big]R(a_+,\beta) \Big\} \\
    - \frac{\Theta(\hbar\omega-2\Delta)}{2} \int_{\Delta-\hbar\omega}^{-\Delta} dE \big[1-2f(E+\hbar\omega)\big] \Big\{ \big[g_1(E)+1\big] R(a_-,\beta) + \big[g_1(E)-1\big]R(a_+,\beta) \Big\} 
    \Bigg\},
    \end{multline}
\end{subequations}
where
\begin{subequations}\label{eps_nb_2}
    \begin{equation}
        f(E) = 1/(1+e^{E/k_bT}), \qquad \beta=1/kl,
    \end{equation}
    \begin{equation}
        S(a,b) = \frac{a}{2} - \frac{ab}{2} \Big[ \arctan{\big(\frac{2b}{b^2+a^2-1}\big)} +\Theta(1-b^2-a^2)\pi \Big] + \frac{1}{8}(1+b^2-a^2)\ln{\Big(\frac{b^2+(1+a)^2}{b^2+(1-1)^2}\Big)},
    \end{equation}
    \begin{equation}
        R(a,b) = -\frac{b}{2} + \frac{ab}{4} \ln{\Big(\frac{b^2+(1+a)^2}{b^2+(1-1)^2}\Big)} + \frac{1}{4}(1+b^2-a^2) \Big[ \arctan{\big(\frac{2b}{b^2+a^2-1}\big)} + \Theta(1-b^2-a^2)\pi \Big],
    \end{equation}
    \begin{equation}
        g_1 = \frac{E^2+\Delta^2+\hbar\omega E}{\sqrt{E^2-\Delta^2}\sqrt{(E+\hbar\omega)^2-\Delta^2}}, \qquad g_2 = \frac{E^2+\Delta^2+\hbar\omega E}{\sqrt{\Delta^2 - E^2}\sqrt{(E+\hbar\omega)^2-\Delta^2}},
    \end{equation}
    \begin{equation}
        a_1 = \frac{\sqrt{E^2-\Delta^2}}{\hbar v_F k}, \qquad a_2 = \frac{\sqrt{(E+\hbar\omega)^2-\Delta^2}}{\hbar v_F k}, \qquad  a_-=a_2-a_1, \qquad a_+=a_1+a_2,
    \end{equation}
\end{subequations}
\end{widetext}
where $l$ is the mean free path, $v_F$ is the Fermi velocity. Equation~(\ref{eps_nb}) reduces to Eq.~(\ref{eps_nb_local}) in the limit $k\ll l^{-1}$. In the paper, we consider the Fermi velocity $v_F\approx 5\times 10^5 \,\mathrm{m/s}$ for niobium~\cite{mattheiss1970electronic} and the mean free path $l=v_F\tau\approx 14.5\,\mathrm{nm}$. We employ Eqs.~(\ref{eps_nb},~\ref{eps_nb_2}) in our calculations in Fig.~\ref{fig:fig5}. In Fig.~\ref{fig:fig6}, we demonstrate the nonlocal transverse permittivity $\varepsilon^\bot_{\mathrm{Nb}}(k,\omega)$ corresponding to different $k$. As shown in Fig.~\ref{fig:fig6}, the nonlocal effects are not appreciable at $k < l^{-1}$. 

\nocite{*}
%\clearpage
\bibliography{reference}% Produces the bibliography via BibTeX.

\end{document}